\documentclass[conference]{IEEEtran}
\makeatletter
\def\ps@headings{%
\def\@oddhead{\mbox{}\scriptsize\rightmark \hfil \thepage}%
\def\@evenhead{\scriptsize\thepage \hfil \leftmark\mbox{}}%
\def\@oddfoot{}%
\def\@evenfoot{}}
\makeatother
\usepackage{subcaption}

\IEEEoverridecommandlockouts

\usepackage{cite}
\usepackage{amsmath,amssymb,amsfonts}
\usepackage{algorithmic}
\usepackage{graphicx}
\usepackage{textcomp}
\usepackage{xcolor}
\usepackage{url}
\usepackage[T1]{fontenc}
\usepackage{flushend}

\def\BibTeX{{\rm B\kern-.05em{\sc i\kern-.025em b}\kern-.08em
    T\kern-.1667em\lower.7ex\hbox{E}\kern-.125emX}}

\renewcommand{\baselinestretch}{1.02}

\begin{document}

\title{DyPBP: Dynamic Peer Beneficialness Prediction for Cryptocurrency P2P Networking}

\author{\IEEEauthorblockN{
        Nazmus Sakib\IEEEauthorrefmark{1},
        Simeon Wuthier\IEEEauthorrefmark{1},
        Amanul Islam\IEEEauthorrefmark{1},
        Xiaobo Zhou\IEEEauthorrefmark{2},
        Jinoh Kim\IEEEauthorrefmark{3},
        Ikkyun Kim\IEEEauthorrefmark{4},
        Sang-Yoon Chang\IEEEauthorrefmark{1}
 }
 \IEEEauthorblockA{\IEEEauthorrefmark{1}\textit{
        Department of Computer Science, 
        University of Colorado, Colorado Springs, CO 80918, USA
 } \\
 \{nsakib, swuthier, aislam2, schang2\}@uccs.edu
        }

 \IEEEauthorblockA{\IEEEauthorrefmark{2}\textit{
        Department of Computer and Information Science, 
        University of Macau, E11 Avenida da Universidade, Taipa, Macau, China
 } \\
 waynexzhou@um.edu.mo
        }
 \IEEEauthorblockA{\IEEEauthorrefmark{3}\textit{
        Department of Computer Science and Information Systems, 
        East Texas A\&M University, Commerce, TX 75428, USA
 } \\
 jinoh.kim@etamu.edu
        }

 \IEEEauthorblockA{\IEEEauthorrefmark{4}\textit{
        Cyber Security Research Division, 
         Electronics and Telecommunications Research Institute, Daejeon, Republic of Korea
 } \\
 ikkim21@etri.re.kr
        }
}

\maketitle

\begin{abstract}
Distributed peer-to-peer (P2P) networking delivers the new blocks and transactions and is critical for the cryptocurrency blockchain system operations. Having poor P2P connectivity reduces the financial rewards from the mining consensus protocol. Previous research defines beneficalness of each Bitcoin peer connection and estimates the beneficialness based on the observations of the blocks and transactions delivery, which are after they are delivered. However, due to the infrequent block arrivals and the sporadic and unstable peer connections, the peers do not stay connected long enough to have the beneficialness score to converge to its expected beneficialness. We design and build Dynamic Peer Beneficialness Prediction (DyPBP) which predicts a peer's beneficialness by using networking behavior observations beyond just the block and transaction arrivals. DyPBP advances the previous research by estimating the beneficialness of a peer connection before it delivers new blocks and transactions. To achieve such goal, DyPBP introduces a new feature for remembrance to address the dynamic connectivity issue, as Bitcoin's peers using distributed networking often disconnect and re-connect. We implement DyPBP on an active Bitcoin node connected to the Mainnet and use machine learning for the beneficialness prediction. Our experimental results validate and evaluate the effectiveness of DyPBP; for example, the error performance improves by 2 to 13 orders of magnitude depending on the machine-learning model selection. DyPBP's use of the remembrance feature also informs our model selection. DyPBP enables the P2P connection's beneficialness estimation from the connection start before a new block arrives. 
\end{abstract}


\section{Introduction} 
\label{sec:intro}
Bitcoin's P2P network enables decentralized financial transactions without the need for centralized authorities. This design ensures that transaction validation and block propagation are handled collectively by a network of miners and nodes, promoting transparency and resistance to censorship. However, the effectiveness of the Bitcoin network depends heavily on the reliability and integrity of its P2P connections. Poor connectivity can reduce the financial rewards earned by incentive-driven nodes through the mining consensus protocol. 
Thus, having peer connections which deliver the blockchain-critical block and transaction information are important. 
We therefore estimate the beneficialness of each peer connection based on the new blocks and transactions they deliver. 

In the existing real-world practice, Bitcoin uses a misbehavior tracking mechanism known as the ban score, which penalizes negative behaviors such as the transmission of invalid messages from peers for tracking its peer connections. While it is designed to improve network security, this mechanism has several limitations. It is vulnerable to identity spoofing and fails to reward positive contributions. 
Furthermore, in the permissionless and pseudonymous environment of Bitcoin, malicious actors can bypass the ban score by switching to new identities, making the system less effective and enabling defamation attacks against innocent peers~\cite{fan2022security}~\cite{fan2021security}.

Previous research to advance the real-world practice designs an alternative 
beneficialness estimation which is based on positive-behavior tracking, rather than penalizing the misbehavior, such as ban score~\cite{wuthier2024positive}. 
The authors 
observe and count the new/previously unseen blocks and transactions to measure the beneficialness~\cite{wuthier2024positive}, which we call the \emph{beneficialness score} ($s$) in this paper. 
However, this approach has limited practicality due to frequent peer disconnections and the sporadic arrival of new blocks. 
For example, blocks are created approximately every 10 minutes in Bitcoin~\cite{nakamoto2008bitcoin}
so, if having 10 connections (i.e., private node), then each peer connection will transmit a new block in every 100 minutes in expectation. 
A typical peer node is connected for a duration less than this, as we will later see in Section~\ref{sec:mot}. 
In addition, many peers do not remain connected long enough for their beneficialness score to converge to the expected beneficialness. 

To overcome these challenges, we design and build a novel, behavior-driven approach for Dynamic Peer Beneficialness Prediction estimation of the cryptocurrency P2P connection. We call our scheme DyPBP which stands for Dynamic Peer Beneficialness Prediction. We utilize dynamic behavioral features
and use
machine learning to estimate peer beneficialness dynamically and before receiving the new blocks.
DyPBP predicts and estimates the beneficialness ($\tilde{s}$) while using the beneficialness score from 
~\cite{wuthier2024positive} as the labels. 

DyPBP also introduces a novel feature for \emph{remembrance} which feature remembers the beneficialness state/score of a previously disconnected peer connection when it gets reconnected. This novel feature has varying impacts on the machine learning models and therefore informs our model selection to have the highest estimation accuracy. 

\label{three-logical-categories}
We implement DyPBP on an active Bitcoin node connected to the Mainnet. 
We sense the Mainnet P2P networking, introduce the remembrance feature, and process the data to implement our DyPBP. 
Our experiment validates our design choices and test the control parameters, including the remembrance feature, the weight control, and machine-learning model selection. 
We measure the accuracy performances to evaluate DyPBP. 


The rest of the paper is organized as follows. We review previous literature in Section~\ref{sec:related}, followed by the background and state-of-the-art limitations in the current Bitcoin P2P network in Section~\ref{sec:back}. Next, in Section~\ref{sec:mot}, we discuss the motivations of our work investigating connectivity behavior.  
We design DyPBP in Section~\ref{sec:proposed_ML}, including how we leverage the previous research by Wuthier et al.~\cite{wuthier2024positive}  
to provide the labels for DyPBP's supervised learning in Section~\ref{sec:beneficialness_score_and_weight}. We describe the data collection and processing of DyPBP in Section~\ref{sec:data_collection} and the experimental results in Section~\ref{sec:results}. We discuss the future work directions in Section~\ref{sec:future} and conclude our work in Section~\ref{sec:conclusion}.


\section{Related Work}
\label{sec:related}

Estimating the beneficialness of peers in cryptocurrency networks helps to maintain secure and efficient P2P operations. Previous works explore beneficialness and connectivity estimation mechanisms for cryptocurrency networks as well as the mechanisms in the general P2P domain. 

\subsection{Peer Beneficialness Estimation in Cryptocurrency}

Our interest in the research of network-based reputation mechanisms in Bitcoin spans from the existing challenges of ban score in Bitcoin's P2P network. The ban score, also known as misbehavior tracking, is a mechanism put in place by the Bitcoin developers as a defense against the denial of service (DoS) threat~\cite{firstBanscoreImplementation,fan2022security}. If a peer connection is caught sending malicious messages to another user (e.g., corrupt packets or invalid cryptographic hashes/digital signatures), then a local counter is maintained, and if this score exceeds an upper threshold, the peer's IP address gets banned for a finite amount of time. 
While the mechanism may help to repel malicious activity, it also makes the host node susceptible to defamation attacks (discussed in Section~\ref{sec:intro}). 
The ban score influences the general motivation to introduce beneficialness as an alternative scoring mechanism. One such mechanism was proposed by Wuthier et al.~\cite{wuthier2024positive}. 
Unlike Bitcoin's traditional ban score penalizing negative peer behavior, their work introduced a beneficialness scoring system that rewarded peers for beneficial actions (specifically the relay of unique blocks and transaction fees). 
Similar works proposed blockchain-based reputation estimation, such as Dennis and Owen~\cite{dennis2015rep} who built a system to store a similar reputation score based on the peer transaction relays. Their work used transaction relays as a means to ensure transparency and consensus of peer reputations across the peer connections. 
Specifically, they incorporated cryptographic proofs for reputation in the transactions, focusing on authenticity and fairness. 

Other research studied the macro-level assessments of connectivity health, i.e., considering the aggregate connection behavior without distinguishing the individual connections. 
Hong et al.~\cite{hong2021robust,hong2022robust} proposed a connectivity estimation engine for the Bitcoin network using statistical analyses. 
Other previous research used
machine learning to connectivity estimation~\cite{kim2021machine} and built anomaly detection~\cite{LION_2023,kim2021anomaly} by drawing insights and patterns from empirical blockchain network traffic. 
However, these previous research assessed general connectivity health by aggregating the effects of multiple peer connections, rather than distinguishing individual peer interactions as we do in this paper.


Although the previously mentioned works focus on connectivity estimation and anomaly detection techniques, it is important to look at the other fields influenced by the information propagation layer of blockchain technologies. One such field includes the modeling of the blockchain network topology of Bitcoin to gain insights orthogonal to beneficialness estimation but with similar techniques. For example, in ~\cite{neudecker2016timing}, Neudecker et al. used the timing of Bitcoin's flooding protocol (specifically INV messages) to observe the timing differences from each of its peer connections. Using these insights, their work aimed to characterize the ``node degree distribution,'' or how connected the peer is to its neighboring peer connections. 
Similarly, Sakib et al.~\cite{sakib2024slow} 
analyzed the impacts of the non-anonymous and anonymous routing protocols supported by Bitcoin (including IP, Tor, I2P, and CJDNS) on its P2P network. 
The goal of this work was to identify and compare the performance of the routing protocols. Their analysis used individual block propagation latencies and throughput to compare networks and emphasizes the need for efficient routing mechanisms to mitigate delays and partitioning. While their work was focused on the generic routing protocols, our work instead addresses peer judging and selection based on the usefulness of peer network behavior.

While it is clear that empirical network-based reputation and beneficialness in cryptocurrency is an active research field, significantly more research works exist beyond the limitations of blockchain-based architectures.

\subsection{Peer Beneficialness Estimation in General P2P Beyond Cryptocurrency}


The beneficialness of users in the networking context is a vast and ever-expanding research domain. An effective estimation engine can predict near-optimal network paths and minimize routing overheads, causing it to be a high priority in some networking applications. One such application is the information sharing field where high availability and short routing paths between users are the priority. 
EigenTrust was introduced by Kamvar et al.~\cite{kamvar2003eigentrust} and represents one of the earliest successful reputation estimation systems designed to mitigate the impact of malicious peers in decentralized networks. By aggregating local trust values assigned by peers to each other, and then normalizing the scores across the network to prevent collusion and other malicious manipulative behaviors, EigenTrust was able to maintain availability and scalability by determining how reliable its neighboring peers are. However, this approach assumes a static environment for trust values, not accounting for dynamic and real-time behavioral changes among peers. To improve the limitations of EigenTrust, GossipTrust was proposed by Zhou et al. to handle real-time updates in trust values dynamically, allowing greater fault tolerance, scalability, and robustness~\cite{zhou2008gossiptrust}. Using a gossip protocol, this system focused on propagating trust information across the network to enable rapid connectivity updates. However, similar to the limitations of EigenTrust, the GossipTrust protocol relied heavily on statistical modeling and predefined heuristics thus limiting its adaptability to fluctuating activity in crowded network environments. 
In another work focusing on fine-grained reputation systems, Xiong et al. introduced a framework for reliable service selection in P2P networks~\cite{zhang2007fine}. This system emphasized the importance of distinguishing between different types of peer behaviors including resource sharing and request handling. 
While these works were being developed, focusing on the optimization of communication through broadcasting and synchronizing trust values of peers, other reputation mechanisms were deviating from performance to focus on the security implications associated with trusting other peers.

Sybilproof and SybilGuard were reputation estimation systems aimed at preventing Sybil attacks by limiting the influence of malicious peers using identity-based constraints and reputation aggregation mechanisms~\cite{cheng2005sybilproof,yu2008sybilguard}. While they proved to be effective in reducing attack surfaces they also often required additional overhead for identity verification, e.g., relying on a trusted central authority.


While the previous studies either focused on sharing trust for network optimization, the attack defenses due to sharing the reputation score publicly, or calculating peer connectivity using static features and utilizing statistical analysis, in our work, we introduce a dynamic, behavior-driven perspective incorporating real-time data and supervised machine learning. Since networking traffic is sequential through time, machine learning-based approaches to beneficialness estimation provide a compelling way to draw insights about empirical network behaviors.


\section{Background and State-of-the-Art Limitations}
\label{sec:back}

\subsection{Bitcoin Peer-to-Peer Network}

\subsubsection{Information Dissemination}
The Bitcoin P2P network facilitates the dissemination of transaction and block data among its nodes to ensure all participants are synchronized with the blockchain. 
Upon creation, transactions are broadcast to peers and stored temporarily in the unconfirmed transaction memory pool (mempool) until a miner inserts it into the next published block. Similarly, upon creation and discovery of a valid block, the data is immediately propagated across the network to reach global acknowledgment as quickly as possible. This is what enables the synchronization and consensus of the blockchain by nodes globally. 

Efficient information dissemination is important for maintaining the integrity and functionality of the Bitcoin network. Delays or disruptions in this process can lead to inconsistencies and consensus issues, such as stale blocks which weaken the fairness and efficiency of mining rewards. In other words, having poor P2P connectivity results in a reduction of the financial rewards by the mining consensus protocol for incentive-driven nodes. Adversarial behaviors such as selective forwarding or intentional propagation delays can disrupt the network, thus highlighting the need for mechanisms that evaluate and encourage proper peer behaviors.

\subsubsection{Permissionless and Trustless Environment}
Bitcoin’s P2P network operates in a permissionless and trustless manner, allowing any participant to join without needing approval or registration. While this openness promotes decentralization, it also introduces vulnerabilities related to identity/trust management. Malicious actors can exploit the system by frequently changing IP addresses to evade detection. This makes traditional identity-based trust systems ineffective in permissionless and trustless environments. 
To address these challenges, dynamic evaluation of peer behaviors provides a way to identify and characterize networking nodes. Instead of relying on static identities, in permissionless networking, the focus shifts to analyzing actions such as transaction and block propagation patterns, connectivity stability, and compliance with protocol rules. This enables trustworthiness to be assessed based on observable behaviors rather than unverifiable claimed identities.

\subsection{Bitcoin P2P Management: Number of Connections and Ban Score}

Bitcoin node has limited number of P2P connections\footnote{There are exceptions for R\&D purposes, involving the source code and implementation are modifications, e.g.,~\cite{wuthier2022greedy}~\cite{chang2025analyzing}.} with varying quality per connection.  
Bitcoin Core restricts private nodes to a maximum of 10 outbound connections. There are small number of public nodes which can additionally have up to 128 inbound connections, as they advertise their IP addresses and accept incoming P2P connection requests.
At the time of writing in December 2024, Bitcoin’s P2P network consists of about 20,000 public nodes, with approximately 92.17\% running Bitcoin Core~\cite{bitnodes}~\cite{sarker2024blockchain}. 
Due to these upper bounds, a node may end up connected to many peers with low connectivity. The low connectivity can delay the relay of blocks and transactions, leading to synchronization issues within the blockchain ledger~\cite{apostolaki2017hijacking}.




Bitcoin implements ban score to keep track of the peer's connectivity (connection quality). Earlier research however demonstrates that Bitcoin’s ban score mechanism is ineffective and vulnerable in real-world practice~\cite{fan2022security}~\cite{fan2021security}. In a permissionless environment, malicious peers can frequently change identities, bypass bans, or spoof genuine peers to trigger their banning. 

The limit in the number of peer connections and the current Bitcoin's ban score limitation motivate our work. Our work aims to build a robust, reliable, and practical mechanism for measuring and estimating peer connection's connectivity. In particular, DyPBP advances the previous research in~\cite{wuthier2024positive} to enhance practicality so that we can preemptively estimate the peer connection's beneficialness before the sporadic block events occur. 


\section{Motivating Our Work: Preliminary Study on Connectivity Behaviors in Bitcoin P2P}
\label{sec:mot}

\begin{figure}[t]
\centering
\includegraphics[width=1\columnwidth]{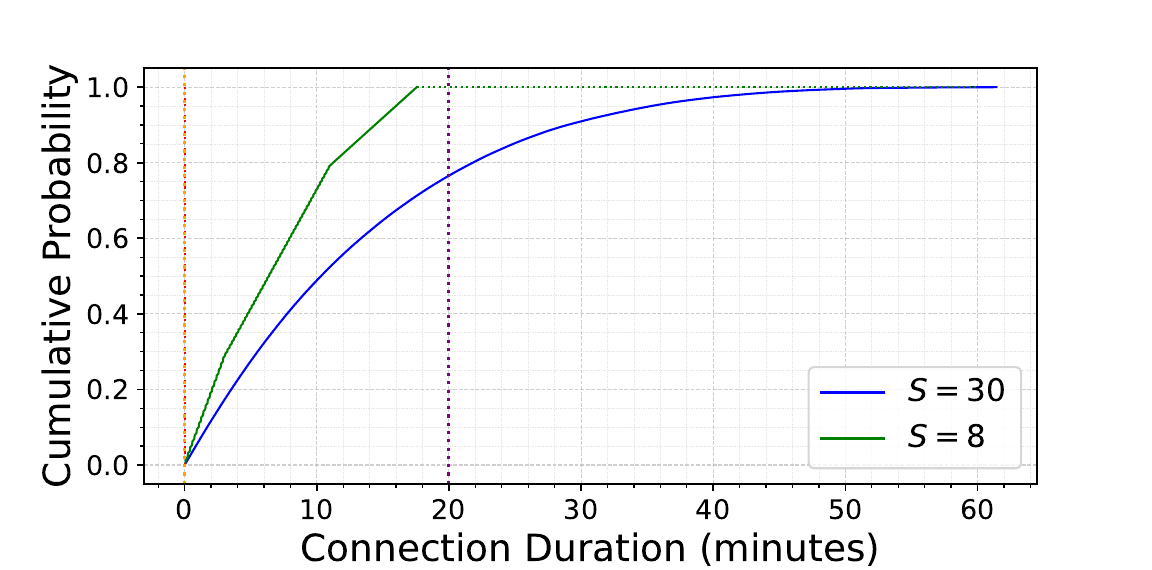}
\caption{CDF of Connection Durations for Different Peer Groups. 
}
\label{fig:CDF_Connection_Duration}
\end{figure}
This section describes our observations of  the Bitcoin P2P networking and the beneficialness scoring in the previous 
work~\cite{wuthier2024positive}. 
The preliminary study and the observations in this section, however, are new (i.e., not in~\cite{wuthier2024positive}) to motivate the advancements in this paper and DyPBP design. 

As discussed in Section~\ref{sec:back}, Bitcoin's P2P networking is dynamic, each peer connection frequently disconnects and re-connects. This is a common behavior in many P2P networking on the Internet, and we observe such behavior in our preliminary study. 
%

Many of the Bitcoin peer connections are not robust and do not sustain long, and the beneficianlness scoring of such peers remain zero as a result.  
Our analysis of approximately 19,500 peers highlights patterns and challenges motivating our research. 
Among these peers, we identify 
approximately 300 peers with near-zero connection duration (depicted as the red curve in Figure~\ref{fig:CDF_Connection_Duration}). These peers fail to maintain any measurable connectivity, resulting in no observable beneficial score. This observation raises an important question: How can we evaluate and account for peers who may contribute briefly or sporadically to the network but are overlooked by traditional beneficialness estimation systems that rely on sustained connection durations? This limitation of current systems became a driving factor for our work.


Furthermore, we observe a clear relationship between connection duration and beneficialness among peers. 
For instance, peers with a low beneficialness score of 8 ($S = 8$), represented by the green curve in Figure~\ref{fig:CDF_Connection_Duration}, are associated with a connection duration of only 18 minutes, indicating less significant contributions to the network. On the other hand, peers with a high beneficialness score of 30 ($S = 30$), represented by the blue curve in Figure~\ref{fig:CDF_Connection_Duration}, exhibit a connection duration of 76 minutes. This positive correlation between prolonged connectivity and beneficial behavior inspires us to explore how these patterns could be systematically integrated into a more advanced and accurate beneficialness estimation solution.

Our calculation of average maximum and minimum connection durations across all peers emphasize the need for an 
a comprehensive analysis. 
The maximum connection duration on average, depicted as the purple curve in Figure~\ref{fig:CDF_Connection_Duration}, is 20 minutes, while the minimum connection duration on average, depicted as the orange curve in Figure~\ref{fig:CDF_Connection_Duration}, is 0.5 seconds. These averages highlight the significant variability in peer connectivity, which simplistic duration-based metrics fail to capture fully. This variability motivates us to design a system that can dynamically evaluate and include peers across the entire spectrum of connectivity behaviors.

The limitations in existing systems, coupled with the observed correlations between connectivity patterns and beneficialness scores, form the foundation of our motivation. While prolonged connectivity often correlates to higher beneficialness, peers with shorter durations or intermittent contributions are frequently overlooked. These insights motivate us to develop a behavior-driven approach for real-time beneficialness estimation, which leverages dynamic metrics to enhance peer evaluation in Bitcoin's P2P network. By addressing these gaps, our work aims to create a more inclusive and accurate solution for assessing peer contributions, improving both the efficiency and trustworthiness of the network.

\section{DyPBP Scheme}
\label{sec:proposed_ML}

DyPBP 
estimates the beneficialness of Bitcoin peers ($\tilde{s}$) by analyzing and leveraging the dynamic, real-time behavioral features of Bitcoin networking traffic. We propose a novel approach that uses networking traffic features rather than relying solely on cryptocurrency-application observations of blocks and transactions. DyPBP utilizes the networking behaviors of the peer connection and applies machine learning to predict how likely they are beneficial in delivering the previously unseen blocks and transactions.

Specifically, we apply supervised machine learning algorithms to estimate the beneficialness of peer nodes based on their observed network behaviors. By deriving the label ($s$) from these observed behaviors, we use machine learning to estimate the beneficialness ($\tilde{s}$) for individual peer nodes.

\subsection{Beneficialness Score and Weight} 
\label{sec:beneficialness_score_and_weight}

We use the beneficialness score from the previous research~\cite{wuthier2024positive} (the previous research calls it the ``reputation score''), which measures the beneficialness using the number of new blocks and transactions. The beneficialness score is for each peer, and the more new block/transactions it delivers, the higher the score. The redundant blocks and transactions (previously received) do not change the score. 
More formally, we define the beneficialness score as $s_t$ where $s$ is the score and it is dependent on $t$, the time duration since the start of logging. $s_t$ is defined by the following.
$$
s_t=s_{t'} + \gamma \sum_{i\in\{B,T\}} w_i f_i
$$
The score, $s_t$, is the sum of the previous score $s_{t'}$ added with a decay factor $\gamma$ and multiplied by the sum of the weighted sensing measurements $w_i f_i$ where $i$ iterates over the sensing parameters from the blocks ($B$) and transactions ($T$). 
The multiplicative decay is a constant value that was introduced in~\cite{wuthier2024positive} to serve as a mechanism preventing peers from accumulating unbounded scores over long durations of time.


The sensing parameters, denoted with $f_i$ measure the empirical count of blocks and the sum of transaction fees received by each peer connection since the past measurement time duration. In other words, the time duration between $t$ and $t'$ contains $f_B$ blocks and $f_T$ satoshi in transaction fees received by the peer connection. 
Because the parameters in $f$ are different units, we introduce weights to prevent the parameters with larger magnitudes from dominating those with smaller magnitudes. To ensure normalized behavior, for all weights, $\sum_i w_i = 1$. The block weight and transaction fee weight (denoted as $w_B$ and $w_T$, respectively) are introduced to balance the ratio of blocks and transaction fees on the computed score, where $w_B+w_T=1$. For example, a larger $w_B$ value causes a larger weight on the block feature while $1-w_B$ is the ratio of weight allocated to the transaction fees ($w_T$). 
If $w_B=0$ (causing $w_T=1$), then $s$ uses and depends only on (previously unseen) transactions. When $w_B=1$, $s$ only uses blocks.
Each peer has its own individual beneficialness score $s, s'$, and network sensing parameters $f$.

\subsection{Remembering the Peer: Remembrance Feature}
\label{sec:remembrance_feature}
Our features measure the networking state, and each measurement is non-overlapping and separate. We introduce a \emph{remembrance} feature to maintain the networking state of a peer between the measurements. The remembrance feature is specific to a peer connection, i.e., every peer has its own remembrance feature. The remembrance feature serves two purposes. First, it connects the current measurement to the previous measurement. Second, if a peer gets disconnected and then re-connected after a period of time, then the remembrance feature remembers the state of the peer before the disconnection. 

We implement remembrance feature by updating the beneficialness score $s$. More specifically, at time $t$, the remembrance feature is $s_{t'}$ where $t'$ is the previous measurement sample time of the peer, e.g., $t-1$ if staying connected. 
If a peer gets disconnected and re-connected, then the remembrance feature enables so that the score $s$ is the score before disconnection. 



The remembrance feature enhances our analysis in several ways. First, it captures the historical peer beneficialness score allowing the machine learning model to consider the impact of past actions. Second, it improves the estimation process by accounting for the gradual increase or decrease in scores over time. This helps machine learning models make more accurate predictions by incorporating patterns of past behavior, thereby enhancing the understanding of beneficialness trends.

\begin{figure}[t]
\includegraphics[width=1\columnwidth]{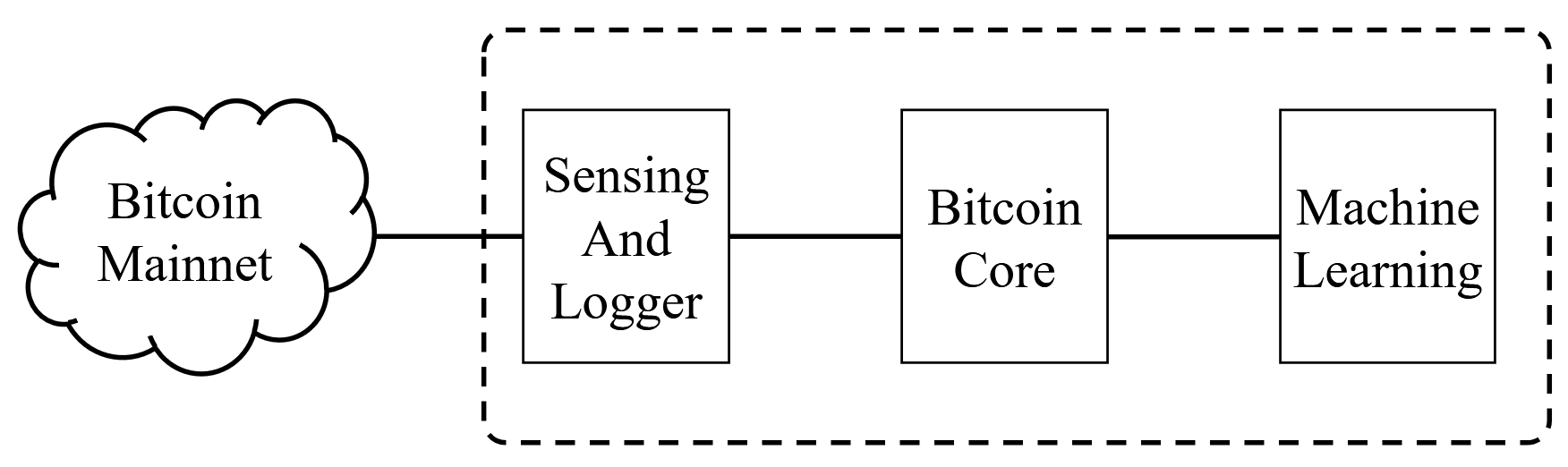}
\caption{Our Implementation from Networking from Bitcoin Mainnet to Sensing to Machine-Learning Processing.}
\label{fig:ML_Model}
\end{figure}

\subsection{From Sensing to ML Processing}
\label{sec:sensing_ml}

Our machine learning implementation is illustrated in Figure~\ref{fig:ML_Model} where we draw the relationship between the steps in our estimation scheme. This consists of the sensing and logging of Bitcoin Core where the data is collected for every peer connection, then the machine learning step that processes the collected data for dynamic real-time beneficialness estimation. 
In the sensing and logger step, the networking messages sent and received for each peer connection are logged into a CSV file for each peer (containing $s$) that we feed into the machine learning algorithms to estimate $\tilde{s}$. 
In other words, supervised machine learning uses the label $s$ and other input data with respect to each peer connection to estimate $\tilde{s}$ such that the error, $|\tilde{s}-s|$ is close to zero. 
This allows our model to learn patterns and relationships between input features and target labels for accurate estimations of future unseen data.

Supervised learning is particularly appropriate for our study as it aligns with the nature of our work in estimating a single behavioral outcome based on several input features. 
We enable the model to generalize and provide accurate estimation for new data by training the model on labeled examples of real-time Bitcoin networking traffic data and their corresponding $s_t$ scores.


In our research, we employ three supervised machine learning models: Linear Regression, Random Forest, and KNN. 
These algorithms are widely used for prediction tasks and are also applied in blockchain networking, e.g.,~\cite{kim2021machine}~\cite{kim2021anomaly}~\cite{harlev2018breaking}~\cite{kim2022machine}. 
We implement these models to compare their prediction performances, including accuracy and to the training duration impacts. The novelty of this work is not in proposing new models but in applying them to peer beneficialness prediction in cryptocurrency P2P networking. As discussed in Section~\ref{sec:beneficialness_score_and_weight}, we use the beneficialness score from prior work~\cite{wuthier2024positive} as the training labels. 

\begin{figure}[t]
\centering
\includegraphics[width=1\columnwidth]{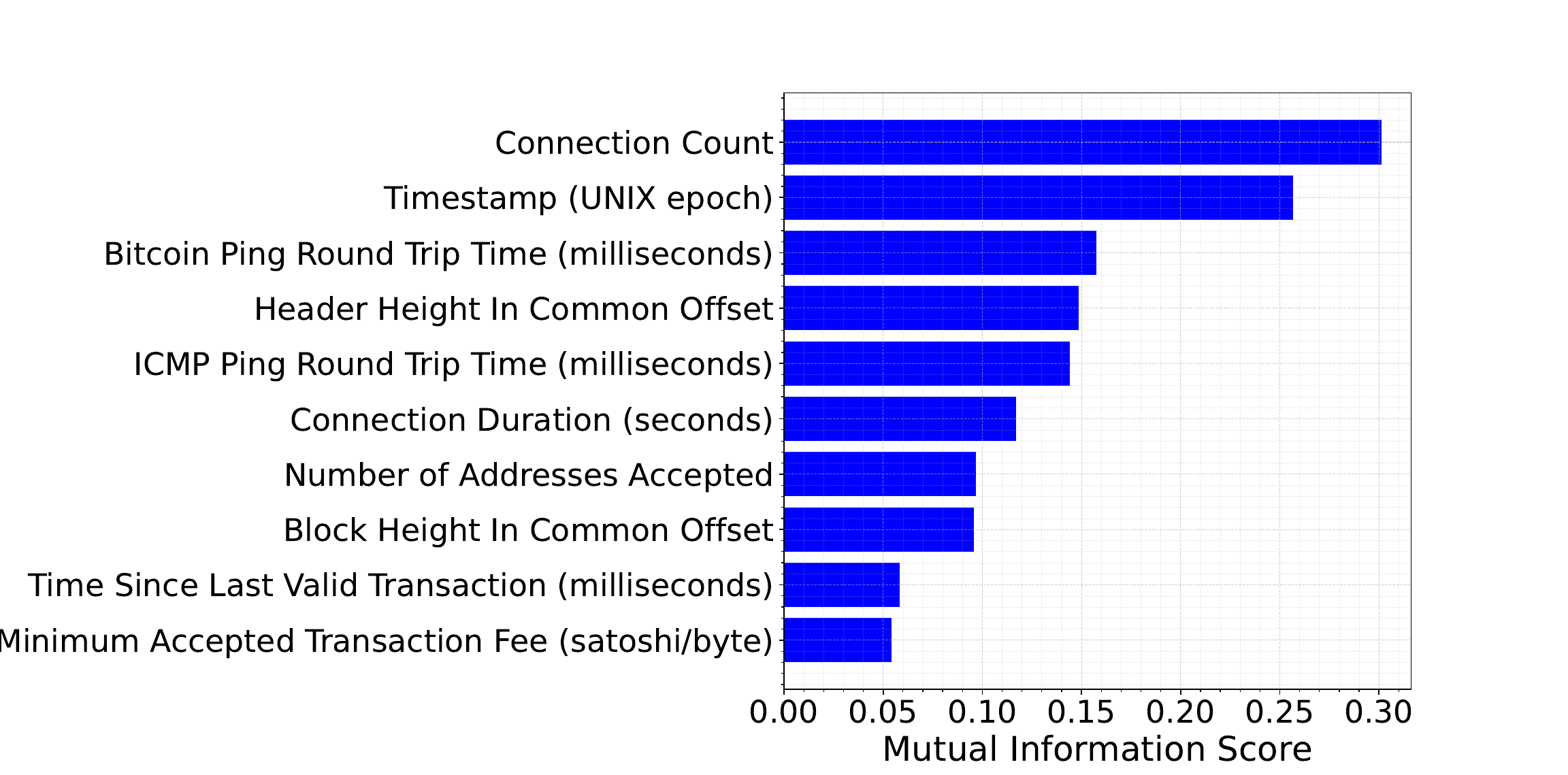}
\caption{Top-10 Features based on Mutual Information.
}
\label{fig:feature_engineering}
\end{figure}

\section{Mainnet Data Collection and Processing}
\label{sec:data_collection}

As mentioned to in Section~\ref{sec:intro}, we implement an active private Bitcoin node prototype connected to the Mainnet. We use Bitcoin Core 0.25.1 and make minimal modifications to link each Bitcoin message with the peer that sent it. 
We collect data for each of the P2P connections, while the number of peers is bounded by ten (i.e., does not exceed ten) because our Bitcoin node is a private node. 
The collected data includes both networking data (retrieved from the Connection Manager and backend Address Manager) and consensus data, which are used to train our machine learning models. 
Our training and testing were conducted on a Linux machine with a 48-core processor at 2.46 GHz and 64 GB of DDR4 RAM.
This section discusses our experimental setup and data collection process.





\subsection{Data Encoding}
In our data from the three logical data collection types (link latency, throughput, and the respective messages from each peer), 272 empirical features are extracted from the networking traffic data. Of these, we identify 80 categorical features. Since categorical data cannot be directly fed into machine learning models we convert them into numerical representations using one-hot encoding~\cite{potdar2017comparative}, a popular encoding technique. 

\subsection{Feature Engineering and Significance}
Feature engineering is essential for 
identifying informative features directly impacting the machine learning model’s ability to assess peer interactions. 
As a preliminary analysis, we identify the ten top influential features for beneficialness estimation, shown in Figure~\ref{fig:feature_engineering}. These include Connection Count, Timestamp (UNIX epoch), Bitcoin Ping Round Trip Time (milliseconds), Header Height In Common Offset, PING Round Trip Time (milliseconds), Connection Duration (seconds), Number of Addresses Accepted, Block Height in Common Offset, Time Since Last Valid Transaction (milliseconds), and Minimum Accepted Transaction Fee (satoshi/bytes). Each of these features captures the behavioral interactions of peers to reflect their impact on connectivity and data transmission. 

Our feature selection process is based on Mutual Information (MI)~\cite{estevez2009normalized}, a useful method for identifying the relevant features in complex datasets. Unlike traditional methods, e.g., correlation that only deals with linear relationships and ANOVA that assumes homogeneity of variance across groups, MI is selected as it can handle non-linear relationships.


\begin{figure*}[t]
\centering
\begin{subfigure}{0.49\linewidth}
    \centering
    \includegraphics[width=1\textwidth]{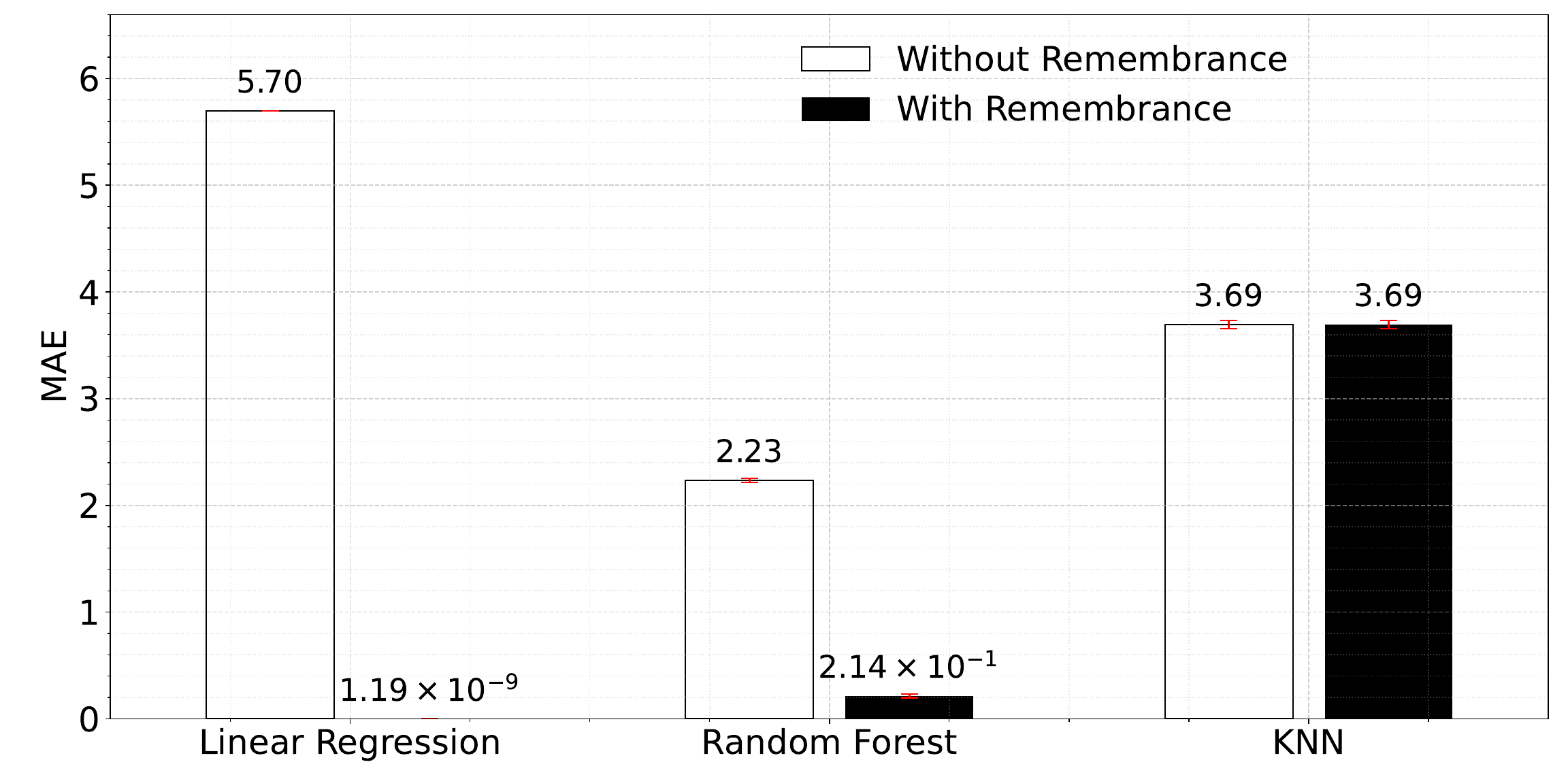}
    \caption{MAE}
    \label{fig:MAE_Compare_LR_RF_KNN}
\end{subfigure}
\hfill
\begin{subfigure}{0.49\linewidth}
    \centering
    \includegraphics[width=1\textwidth]{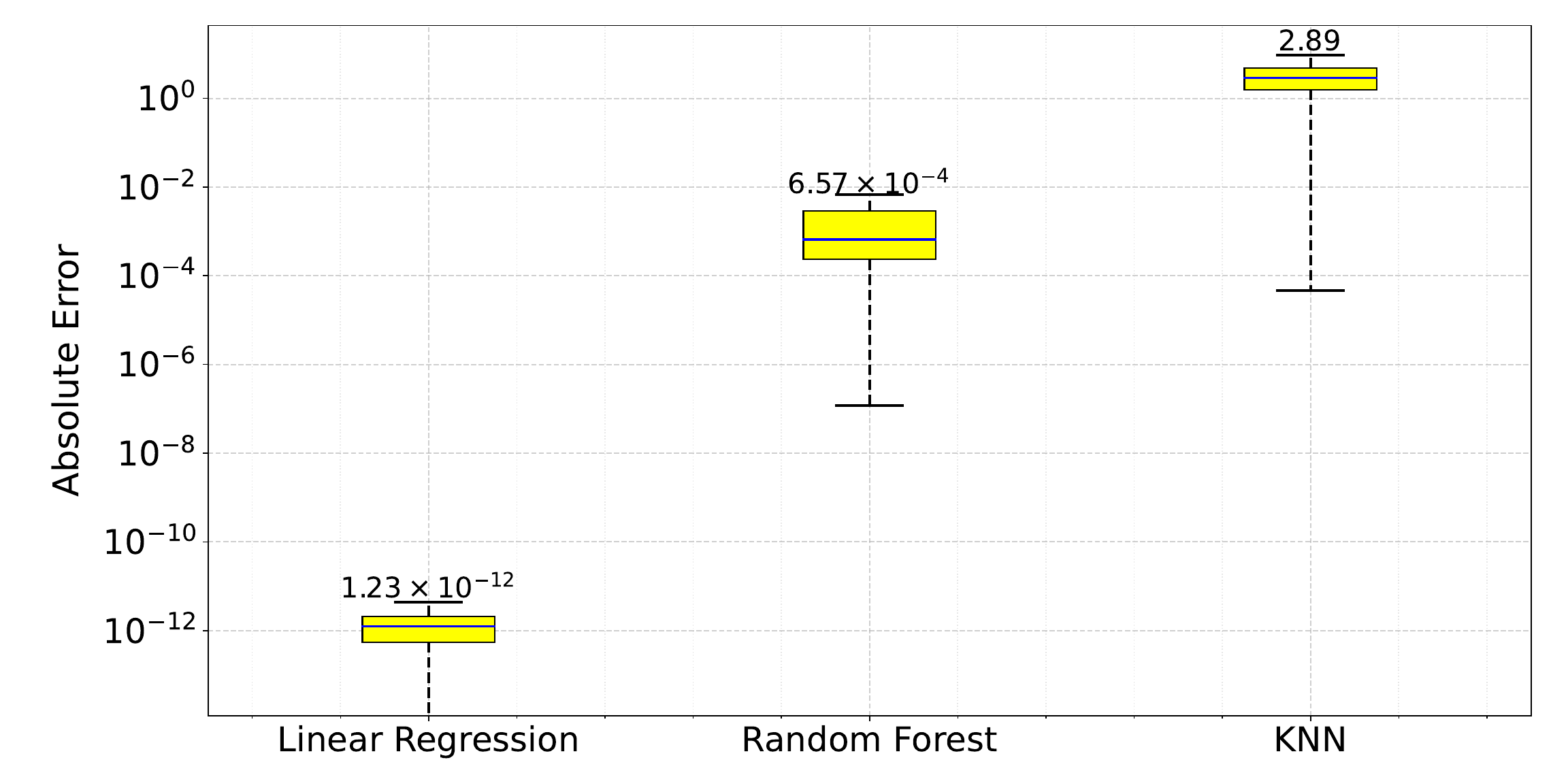}
    \caption{Error Distribution}
    \label{fig:Error_Distribution_Compare_LR_RF_KNN}
\end{subfigure}
\caption{Model Performance Across Machine Learning Algorithms when Fixing $w_B=0.5$ and With vs. Without Remembrance. 
}
\end{figure*}

\section{Experimental Results}
\label{sec:results}


\subsection{Measurements and Control Parameters}





In this project, the mean absolute error (MAE) is used as the performance metric to evaluate the accuracy of our machine learning models in estimating the beneficialness ($s_t$) for each Bitcoin peer. The higher the MAE the less accurate the prediction estimation is. MAE is particularly relevant as it provides an intuitive measure of prediction errors by reflecting the average 
deviation of estimated beneficialness scores ($\tilde{s}$) from their actual values ($s$), allowing us to predict the effectiveness of our models in a dynamic and decentralized environment.
Each peer connection has its beneficial score estimation $\tilde{s}$, and MAE = E[$s-\tilde{s}$] where E[.] is the expected value averaged over all peer connections. 

For the errors, beyond the average/mean in MAE, we additionally present the distribution of prediction errors using box plots and 95\% confidence intervals. The box plots illustrate how individual errors vary across models, highlighting their spread, while the confidence intervals quantify the range within which the true mean error is expected to lie.

As discussed in Section~\ref{sec:remembrance_feature}, remembrance is incorporated through a historical previous score to allow the training process to know the prior state of the score values. Using the weights described in Section~\ref{sec:beneficialness_score_and_weight}, we calculate the score over several ratios of blocks and transaction fee weights where $w_B \in \{0.00, 0.25, 0.50, 0.75, 1.00\}$. 
We compare two versions, \emph{with} and \emph{without} remembrance to study the effect of stateful and stateless training, respectively. Training with remembrance allows the model to recognize patterns over time as opposed to without remembrance.

\subsection{Model Evaluation}
\label{sec:MAE_Results}

\subsubsection{Comparison of MAE Across Machine Learning Algorithms}
\label{sec:comparison_algs}
For our preliminary experimentation, we train three different machine learning algorithms discussed in Section~\ref{sec:proposed_ML} to study the impact that remembrance has on their performance when the $w_B$ is fixed to 0.5. According to Figure~\ref{fig:MAE_Compare_LR_RF_KNN}, Linear Regression and Random Forest show a significant improvement (lower MAE) when remembrance is used (9, and 1 orders of magnitude lower MAE for Linear Regression and Random Forest, respectively). However, the KNN model shows no improvement when comparing MAE values with and without remembrance (where the MAE is fixed at 3.69). This is because KNN generally does not work well with features of different scales, e.g., categorical and non-normalized features which are present within our dataset. Due to this, we largely omit KNN from our remaining experimentation and discussion.

\subsubsection{Beyond MAE: Comparison of Error Distributions Across Models}
\label{sec:comparison_algs_errors_distributions}

While MAE shows the average prediction error, it does not reveal how errors are distributed across test samples. Figure~\ref{fig:Error_Distribution_Compare_LR_RF_KNN} demonstrates the distribution of absolute errors for each model with $w_B$ fixed at 0.5 and the remembrance feature enabled. Linear Regression achieves the most stable performance, with a median error of $1.23 \times 10^{-12}$ and a very narrow interquartile range ($5.4 \times 10^{-13}$ to $2.1 \times 10^{-12}$), showing that most predictions remain extremely close to the ground truth. Random Forest attains a median error of $6.6 \times 10^{-4}$ with most errors below $10^{-2}$. By contrast, KNN produces a much higher median error ($\approx 2.89$) with a wide spread (1.56 to 4.75), confirming its poor suitability for our dataset.





\begin{figure*}[t]
\centering
\begin{subfigure}{0.49\linewidth}
    \centering
    \includegraphics[width=1\textwidth]{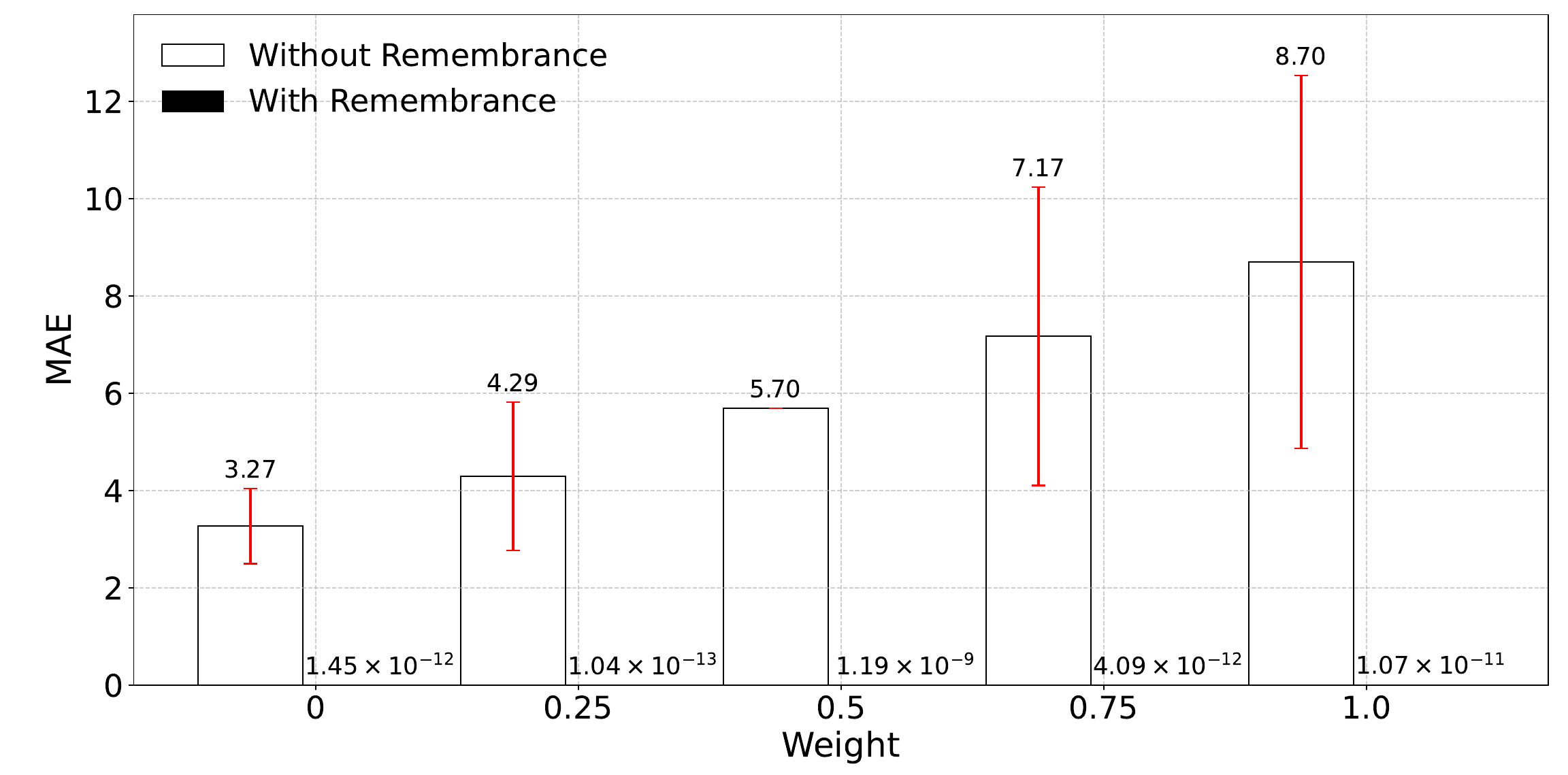}
    \caption{Linear Regression}
    \label{fig:MAE_Compare_LR}
\end{subfigure}
\hfill
\begin{subfigure}{0.49\linewidth}
    \centering
    \includegraphics[width=1\textwidth]{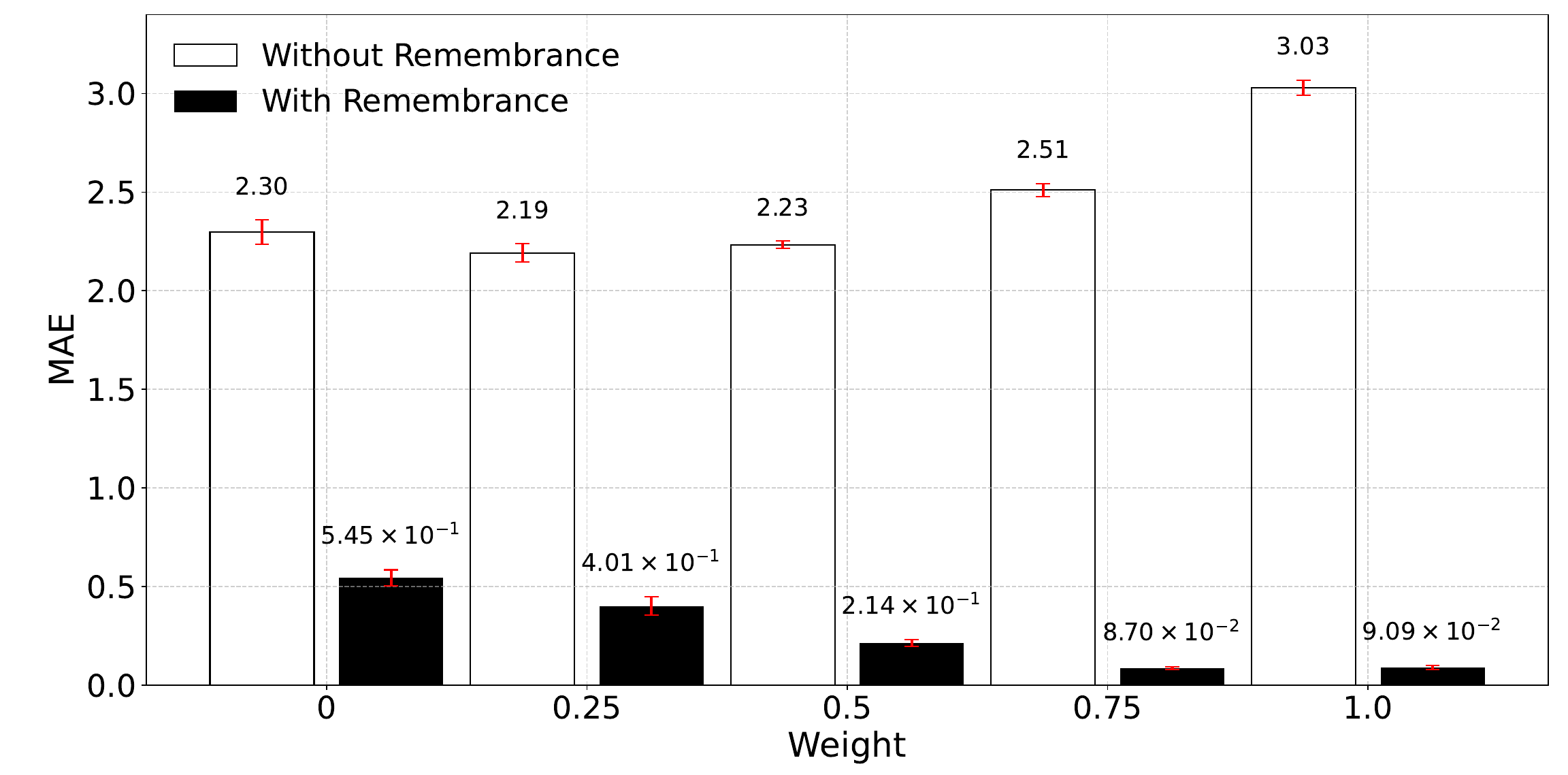}
    \caption{Random Forest}
    \label{fig:MAE_Compare_RF}
\end{subfigure}
\caption{Model Performance of Linear Regression and Random Forest when Varying Weights and if Remembrance is on.}
\end{figure*}

\begin{figure*}[t]
\centering
\begin{subfigure}{0.32\linewidth}
    \centering
    \includegraphics[width=1\textwidth]{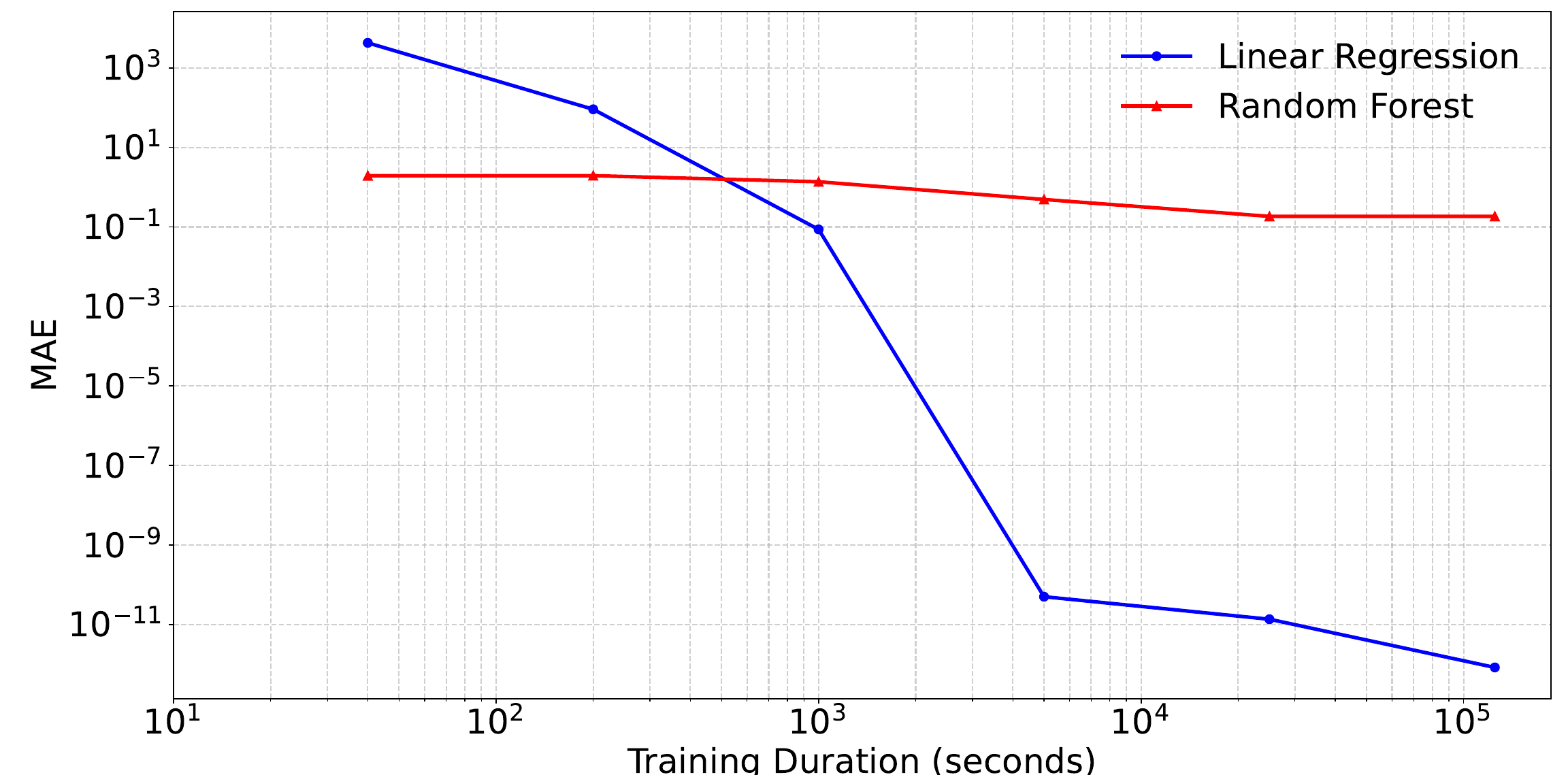}
    \caption{Two Models when $w_B=0.50$}
    \label{fig:MAE_Varying_Training_Duration_LR_RF_0.5}
\end{subfigure}
\hfill
\begin{subfigure}{0.32\linewidth}
    \centering
    \includegraphics[width=1\textwidth]{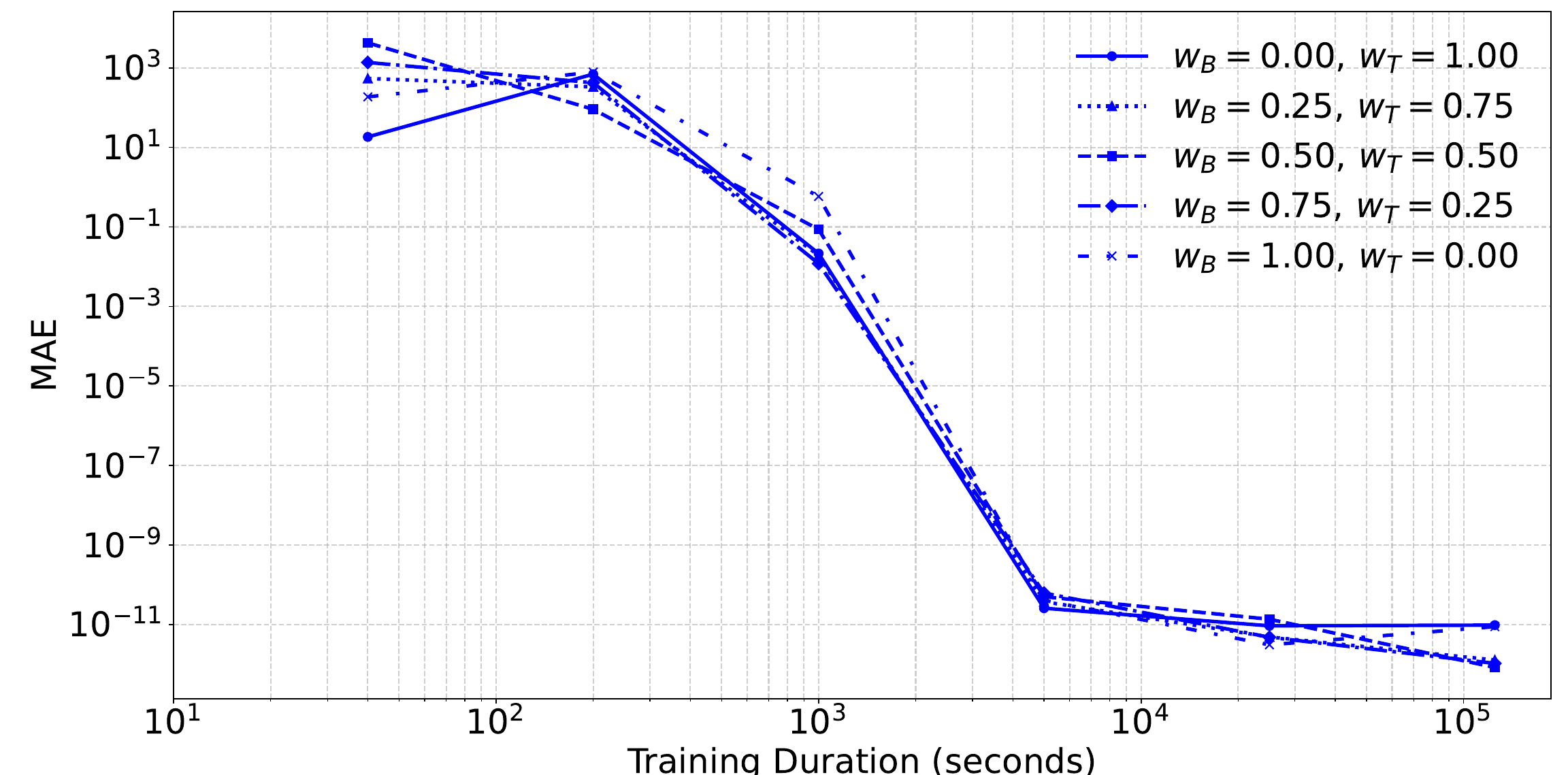}
    \caption{\vspace{3px}Linear Regression when Varying $w_B$}
    \label{fig:MAE_Varying_Training_Duration_LR}
\end{subfigure}
\hfill
\begin{subfigure}{0.32\linewidth}
    \centering
    \includegraphics[width=1\textwidth]{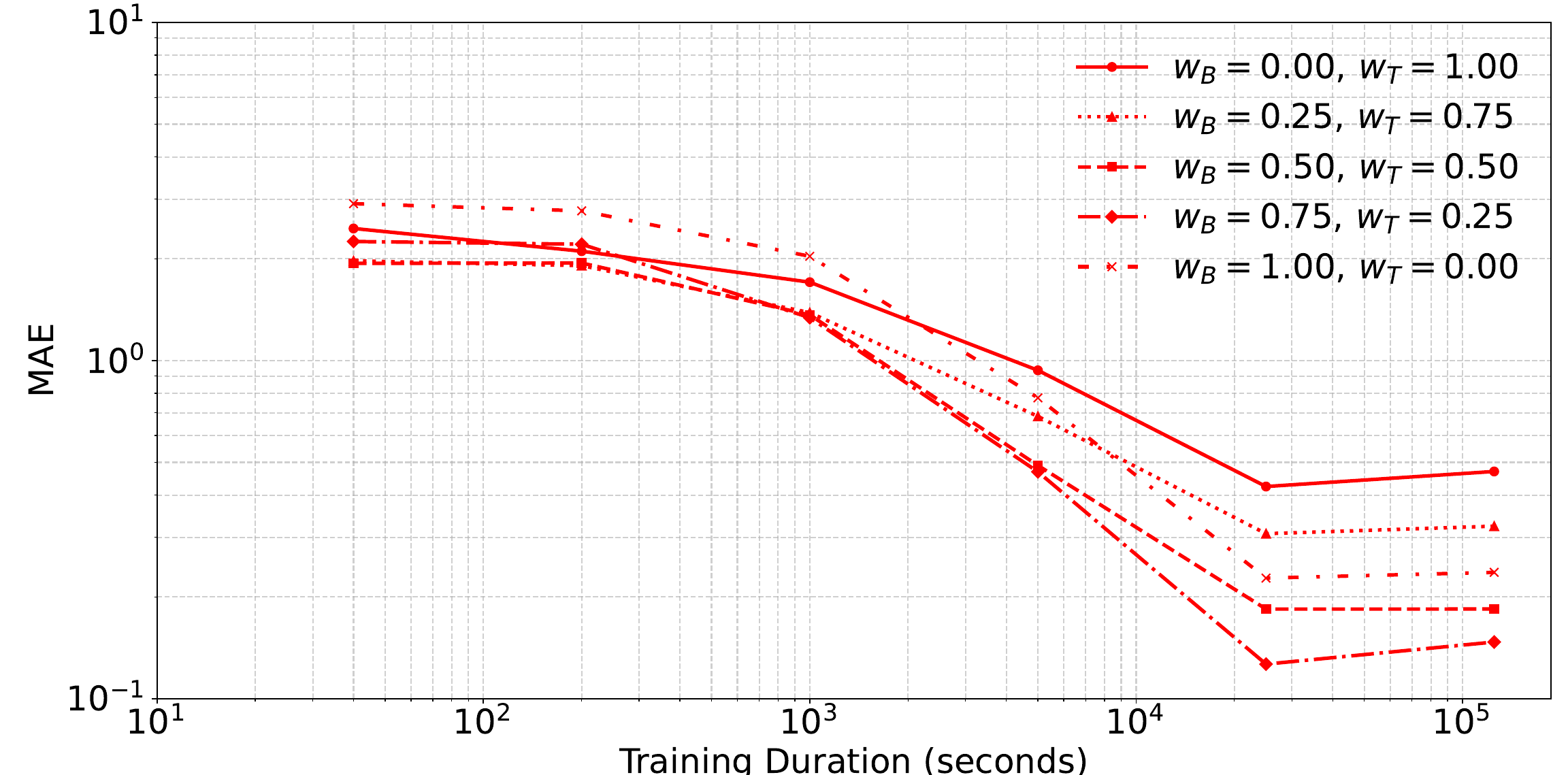}
    \caption{\vspace{3px}Random Forest when Varying $w_B$}
    \label{fig:MAE_Varying_Training_Duration_RF}
\end{subfigure}
\caption{Model Performance of Linear Regression and Random Forest when Varying $w_B$ and Training Durations.}
\end{figure*}

\subsubsection{Comparison of MAE With Varying Weights}
%
%
The results in Figure~\ref{fig:MAE_Compare_LR} and Figure~\ref{fig:MAE_Compare_RF} show clear differences in MAE when toggling remembrance, and when shifting the ratio of $w_B$ and $w_T$ between blocks and transaction fees. Both Linear Regression and Random Forest perform well when the remembrance is included, with the MAE being consistently lower.

On the horizontal access of Figure~\ref{fig:MAE_Compare_LR}, $w_B$ shifts from 0 to 1 where the Linear Regression model shows a much lower MAE when remembrance is included, even reaching $1.04 \times 10^{-13}$ while the MAE without remembrance was 13 orders of magnitude larger for $w_B=0.25$. In general, not including remembrance yields a significantly larger MAE with the largest value being $8.70$ at $w_B=1.00$.
Similarly, for Random Forest in Figure~\ref{fig:MAE_Compare_RF}, the behavior persists where remembrance yields a significantly lower MAE value, even reaching $8.70\times 10^{-2}$ while the MAE without remembrance is 2 orders of magnitude larger for $w_B=0.75$. Since the effect of removing remembrance is less pronounced than Linear Regression, this indicates that Random Forest can perform reasonably well without the historical context. 

As the weight ratio shifts towards a larger dependence on the transaction fees (i.e., $w_B\rightarrow0$), we notice the monotonic decrease in MAE when remembrance is disabled for Linear Regression and a similar behavior for Random Forest except when $w_B=0$. The general decrease in MAE as $w_B$ decreases is due to the difference in frequency of receiving blocks compared to transactions. Since blocks are sporadic (only arriving approximately once every 10 minutes), there is significantly less data than transactions where Bitcoin receives approximately 3-7 transactions per second~\cite{blockchainBlockchaincomCharts}. 
In both cases, including remembrance improves the score estimation performance indicating that the remembrance feature helps the models in identifying the patterns and behaviors that are dependent over time.



\subsubsection{Comparison of MAE With Varying Training Durations}
We study how the training duration impacts the stability and accuracy of the results over different lengths of training data, with the purpose being to determine how much training time each model needs to estimate the score reliably. As shown in Figure~\ref{fig:MAE_Varying_Training_Duration_LR_RF_0.5}, both models show a decrease in MAE as the training duration increases. This indicates that more training data helps each model learn and generalize better, but it is a diminishing return as the improvements in accuracy become less significant with each increment of training duration.

For Linear Regression, MAE significantly decreases with training durations up to 5000 seconds, after which it diminishes (from around 4246 to $8\times 10^{-13}$ for 40 seconds and 125,000 seconds, respectively). This shows that Linear Regression, being a simpler model, reaches stable and lower MAE values relatively quickly to the linear relationships between the features in our dataset. On the other hand, Random Forest, which excels in capturing non-linear relationships, decreases more gradually, limiting its performance (from around $1.92$ to $1.84\times 10^{-1}$ for 40 seconds and 125,000 seconds, respectively). 

Focusing on Linear Regression we show the impact on model training performance as $w_B$ interpolates from 0 to 1 in Figure~\ref{fig:MAE_Varying_Training_Duration_LR}. Results show that balanced configurations around $w_B=0.50$ achieve lower MAE values early on, indicating the model’s efficiency in simpler scenarios. In contrast, when the weights favor one feature (e.g., $w_B=0$ and $w_B=1$), MAE values are larger. Therefore balancing both $w_B$ and $w_T$ generally performs better with larger training durations. 

Similarly, we study the impact of varying $w_B$ on the Random Forest model training in Figure~\ref{fig:MAE_Varying_Training_Duration_RF}. Similar to Linear Regression, Random Forest performs best when $0.25\geq w\geq 0.5$ for smaller training durations (less than 1000 seconds). For greater training durations, $w_B=0.75$ performs best. Similarly, Random Forest performs worst for $w_B=1$ when the training durations are less than approximately 2000 seconds, and when training durations are greater, $w_B=0.75$ performs worst.

\section{Future Work Discussions}
\label{sec:future}
We outline future directions building on this paper 
to facilitate such future R\&D to advance the P2P networking for cryptocurrency blockchain.  

\subsubsection{Active Control Implementation}
The Bitcoin network currently relies on a probabilistic peer selection mechanism, which does not dynamically account for peer behaviors. A promising direction for future work is the implementation of active control mechanisms. By integrating our machine learning-based beneficialness estimation engine into an active Bitcoin node, we aim to create a dynamic peer selection system. This solution would enable nodes to selectively establish outbound connections with the most beneficial peers based on their estimated beneficialness scores. By replacing the existing probabilistic mechanism with this intelligent alternative, DyPBP can enhance the reliability of the network while improving its efficiency and security. Such an approach would allow Bitcoin nodes to adapt to real-time conditions and prioritize beneficial peer connections, strengthening the overall robustness of the P2P network.

\subsubsection{Identification of Malicious Peers}
Another critical area for future exploration is the identification of malicious peers within the Bitcoin network. Using our behavior-driven dataset, we plan to apply machine learning techniques to detect and classify peers engaging in malicious activities, such as spoofing, selective message forwarding, or other disruptive behaviors. This capability would provide nodes with the ability to proactively identify and mitigate risks posed by such adversaries. By leveraging the dynamic behavioral features included in our dataset, we aim to develop a scheme for distinguishing between benign and malicious peers. This work has the potential to significantly enhance the network’s resilience against adversarial threats and ensure a safer environment for transaction propagation and consensus processes.


\section{Conclusion}
\label{sec:conclusion}
Bitcoin’s P2P network relies on effective peer connectivity for efficient block and transaction propagation. Existing beneficialness estimation struggles with frequent peer disconnections and sporadic block arrivals. To address these challenges, we introduce a dynamic, real-time beneficialness estimation scheme using supervised machine learning. Our findings show that adding a remembrance feature to the beneficialness prediction model significantly improves performance for both Random Forest and Linear Regression, reducing the MAE considerably. Furthermore, we observe that training beyond approximately 5000 seconds yields diminishing returns, indicating that extended training durations do not proportionately improve accuracy. These improvements have direct implications for the operational efficiency and security of Bitcoin's P2P network. By enabling timely predictions of peer beneficialness, DyPBP aligns with the mining incentives 
and
advances the active P2P control for cryptocurrency blockchain. 

\bibliographystyle{IEEEtran}

\begin{thebibliography}{10}
\providecommand{\url}[1]{#1}
\csname url@samestyle\endcsname
\providecommand{\newblock}{\relax}
\providecommand{\bibinfo}[2]{#2}
\providecommand{\BIBentrySTDinterwordspacing}{\spaceskip=0pt\relax}
\providecommand{\BIBentryALTinterwordstretchfactor}{4}
\providecommand{\BIBentryALTinterwordspacing}{\spaceskip=\fontdimen2\font plus
\BIBentryALTinterwordstretchfactor\fontdimen3\font minus \fontdimen4\font\relax}
\providecommand{\BIBforeignlanguage}[2]{{%
\expandafter\ifx\csname l@#1\endcsname\relax
\typeout{** WARNING: IEEEtran.bst: No hyphenation pattern has been}%
\typeout{** loaded for the language `#1'. Using the pattern for}%
\typeout{** the default language instead.}%
\else
\language=\csname l@#1\endcsname
\fi
#2}}
\providecommand{\BIBdecl}{\relax}
\BIBdecl

\bibitem{fan2022security}
W.~Fan, S.~Wuthier, H.-J. Hong, X.~Zhou, Y.~Bai, and S.-Y. Chang, ``The security investigation of ban score and misbehavior tracking in bitcoin network,'' in \emph{2022 IEEE 42nd International Conference on Distributed Computing Systems (ICDCS)}.\hskip 1em plus 0.5em minus 0.4em\relax IEEE, 2022, pp. 191--201.

\bibitem{fan2021security}
W.~Fan, H.-J. Hong, S.~Wuthier, X.~Zhou, Y.~Bai, and S.-Y. Chang, ``Security analyses of misbehavior tracking in bitcoin network,'' in \emph{2021 IEEE International Conference on Blockchain and Cryptocurrency (ICBC)}.\hskip 1em plus 0.5em minus 0.4em\relax IEEE, 2021, pp. 1--3.

\bibitem{wuthier2024positive}
S.~Wuthier, N.~Sakib, and S.-Y. Chang, ``Positive reputation score for bitcoin p2p network,'' in \emph{2024 IEEE 21st Consumer Communications \& Networking Conference (CCNC)}.\hskip 1em plus 0.5em minus 0.4em\relax IEEE, 2024, pp. 519--524.

\bibitem{nakamoto2008bitcoin}
S.~Nakamoto, ``Bitcoin: A peer-to-peer electronic cash system,'' \emph{Satoshi Nakamoto}, 2008.

\bibitem{firstBanscoreImplementation}
\BIBentryALTinterwordspacing
Bitcoin, ``Merge pull request \#517 from gavinandresen/dosprevention.'' [Online]. Available: \url{https://github.com/bitcoin/bitcoin/commit/17e2c24645a10354849dec917b31f364e9056d58}
\BIBentrySTDinterwordspacing

\bibitem{dennis2015rep}
R.~Dennis and G.~Owen, ``Rep on the block: A next generation reputation system based on the blockchain,'' in \emph{2015 10th International Conference for Internet Technology and Secured Transactions (ICITST)}.\hskip 1em plus 0.5em minus 0.4em\relax IEEE, 2015, pp. 131--138.

\bibitem{hong2021robust}
H.-J. Hong, W.~Fan, S.~Wuthier, J.~Kim, X.~Zhou, C.~E. Chow, and S.-Y. Chang, ``Robust p2p connectivity estimation for permissionless bitcoin network,'' in \emph{2021 IEEE/ACM 29th International Symposium on Quality of Service (IWQOS)}.\hskip 1em plus 0.5em minus 0.4em\relax IEEE, 2021, pp. 1--6.

\bibitem{hong2022robust}
H.-J. Hong, W.~Fan, S.~Wuthier, J.~Kim, C.~E. Chow, X.~Zhou, and S.-Y. Chang, ``Robust p2p networking connectivity estimation engine for permissionless bitcoin cryptocurrency,'' \emph{Computer Networks}, vol. 219, p. 109436, 2022.

\bibitem{kim2021machine}
J.~Kim, M.~Nakashima, W.~Fan, S.~Wuthier, X.~Zhou, I.~Kim, and S.-Y. Chang, ``A machine learning approach to peer connectivity estimation for reliable blockchain networking,'' in \emph{2021 IEEE 46th Conference on Local Computer Networks (LCN)}.\hskip 1em plus 0.5em minus 0.4em\relax IEEE, 2021, pp. 319--322.

\bibitem{LION_2023}
W.~Fan, H.-J. Hong, J.~Kim, S.~Wuthier, M.~Nakashima, X.~Zhou, C.-H. Chow, and S.-Y. Chang, ``Lightweight and identifier-oblivious engine for cryptocurrency networking anomaly detection,'' \emph{IEEE Transactions on Dependable and Secure Computing}, vol.~20, no.~2, pp. 1302--1318, 2023.

\bibitem{kim2021anomaly}
J.~Kim, M.~Nakashima, W.~Fan, S.~Wuthier, X.~Zhou, I.~Kim, and S.-Y. Chang, ``Anomaly detection based on traffic monitoring for secure blockchain networking,'' in \emph{2021 IEEE International Conference on Blockchain and Cryptocurrency (ICBC)}.\hskip 1em plus 0.5em minus 0.4em\relax IEEE, 2021, pp. 1--9.

\bibitem{neudecker2016timing}
T.~Neudecker, P.~Andelfinger, and H.~Hartenstein, ``Timing analysis for inferring the topology of the bitcoin peer-to-peer network,'' in \emph{2016 Intl IEEE Conferences on Ubiquitous Intelligence \& Computing, Advanced and Trusted Computing, Scalable Computing and Communications, Cloud and Big Data Computing, Internet of People, and Smart World Congress (UIC/ATC/ScalCom/CBDCom/IoP/SmartWorld)}.\hskip 1em plus 0.5em minus 0.4em\relax IEEE, 2016, pp. 358--367.

\bibitem{sakib2024slow}
N.~Sakib, S.~Wuthier, K.~Zhang, X.~Zhou, and S.-Y. Chang, ``From slow propagation to partition: Analyzing bitcoin over anonymous routing,'' in \emph{2024 IEEE International Conference on Blockchain and Cryptocurrency (ICBC)}.\hskip 1em plus 0.5em minus 0.4em\relax IEEE, 2024, pp. 377--385.

\bibitem{kamvar2003eigentrust}
S.~D. Kamvar, M.~T. Schlosser, and H.~Garcia-Molina, ``The eigentrust algorithm for reputation management in p2p networks,'' in \emph{Proceedings of the 12th international conference on World Wide Web}, 2003, pp. 640--651.

\bibitem{zhou2008gossiptrust}
R.~Zhou, K.~Hwang, and M.~Cai, ``Gossiptrust for fast reputation aggregation in peer-to-peer networks,'' \emph{IEEE transactions on knowledge and data engineering}, vol.~20, no.~9, pp. 1282--1295, 2008.

\bibitem{zhang2007fine}
Y.~Zhang and Y.~Fang, ``A fine-grained reputation system for reliable service selection in peer-to-peer networks,'' \emph{IEEE Transactions on Parallel and Distributed Systems}, vol.~18, no.~8, pp. 1134--1145, 2007.

\bibitem{cheng2005sybilproof}
A.~Cheng and E.~Friedman, ``Sybilproof reputation mechanisms,'' in \emph{Proceedings of the 2005 ACM SIGCOMM workshop on Economics of peer-to-peer systems}, 2005, pp. 128--132.

\bibitem{yu2008sybilguard}
H.~Yu, M.~Kaminsky, P.~B. Gibbons, and A.~D. Flaxman, ``Sybilguard: defending against sybil attacks via social networks,'' \emph{IEEE/ACM Transactions on networking}, vol.~16, no.~3, pp. 576--589, 2008.

\bibitem{wuthier2022greedy}
S.~Wuthier, P.~Chandramouli, X.~Zhou, and S.-Y. Chang, ``Greedy networking in cryptocurrency blockchain,'' in \emph{IFIP International Conference on ICT Systems Security and Privacy Protection}.\hskip 1em plus 0.5em minus 0.4em\relax Springer, 2022, pp. 343--359.

\bibitem{chang2025analyzing}
S.-Y. Chang, N.~Sakib, S.~Wuthier, and K.~Paarporn, ``Analyzing and modeling connection impact on distributed consensus in cryptocurrency blockchain,'' in \emph{NOMS 2025-2025 IEEE Network Operations and Management Symposium}.\hskip 1em plus 0.5em minus 0.4em\relax IEEE, 2025, pp. 1--5.

\bibitem{bitnodes}
\BIBentryALTinterwordspacing
``Bitcoin peer-to-peer network, 2024,'' accessed: December, 2024. [Online]. Available: \url{https://bitnodes.io/}
\BIBentrySTDinterwordspacing

\bibitem{sarker2024blockchain}
A.~Sarker, S.~Wuthier, J.~Kim, J.~Kim, and S.-Y. Chang, ``Blockchain handshaking with software assurance: Version++ protocol for bitcoin cryptocurrency,'' \emph{Electronics}, vol.~13, no.~19, p. 3857, 2024.

\bibitem{apostolaki2017hijacking}
M.~Apostolaki, A.~Zohar, and L.~Vanbever, ``Hijacking bitcoin: Routing attacks on cryptocurrencies,'' in \emph{2017 IEEE symposium on security and privacy (SP)}.\hskip 1em plus 0.5em minus 0.4em\relax IEEE, 2017, pp. 375--392.

\bibitem{harlev2018breaking}
M.~A. Harlev, H.~Sun~Yin, K.~C. Langenheldt, R.~Mukkamala, and R.~Vatrapu, ``Breaking bad: De-anonymising entity types on the bitcoin blockchain using supervised machine learning,'' 2018.

\bibitem{kim2022machine}
J.~Kim, M.~Nakashima, W.~Fan, S.~Wuthier, X.~Zhou, I.~Kim, and S.-Y. Chang, ``A machine learning approach to anomaly detection based on traffic monitoring for secure blockchain networking,'' \emph{IEEE Transactions on Network and Service Management}, vol.~19, no.~3, pp. 3619--3632, 2022.

\bibitem{potdar2017comparative}
K.~Potdar, T.~S. Pardawala, and C.~D. Pai, ``A comparative study of categorical variable encoding techniques for neural network classifiers,'' \emph{International journal of computer applications}, vol. 175, no.~4, pp. 7--9, 2017.

\bibitem{estevez2009normalized}
P.~A. Est{\'e}vez, M.~Tesmer, C.~A. Perez, and J.~M. Zurada, ``Normalized mutual information feature selection,'' \emph{IEEE Transactions on neural networks}, vol.~20, no.~2, pp. 189--201, 2009.

\bibitem{blockchainBlockchaincomCharts}
``{B}lockchain.com | {C}harts - {T}ransaction {R}ate {P}er {S}econd --- blockchain.com,'' \url{https://www.blockchain.com/explorer/charts/transactions-per-second}, [Accessed 11-01-2025].

\end{thebibliography}

\end{document}